\documentclass[pre,amsmath,amssymb,floatfix,twocolumn,showpacs]{revtex4}
\usepackage{graphicx}
\usepackage{float}
\usepackage{bm}
\usepackage[utf8]{inputenc}

\begin{document}
\title{The criticality of self-assembled rigid rods on triangular lattices}
\author{N. G. Almarza}
\affiliation{Instituto de Qu{\'\i}mica F{\'\i}sica Rocasolano, CSIC, Serrano 119, E-28006 Madrid, Spain }
\author{J. M. Tavares}
\affiliation{Centro de F\'{\i}sica Te\'orica e Computacional, Universidade de Lisboa, Avenida Professor Gama Pinto 2,
P-1649-003 Lisbon, Portugal}
\affiliation{Instituto Superior de Engenharia de Lisboa, Rua Conselheiro Em\'{\i}dio Navarro 1, 
P-1950-062 Lisbon, Portugal}
\author{M. M. Telo da Gama}
\affiliation{Centro de F\'{\i}sica Te\'orica e Computacional, Universidade de Lisboa, Avenida Professor Gama Pinto 2,
P-1649-003 Lisbon, Portugal}
\affiliation{Departamento de F\'{\i}sica, Faculdade de Ci\^encias, Universidade de Lisboa, Campo Grande,
P-1749-016 Lisbon, Portugal}

\date{\today}
\begin{abstract}
The criticality of self-assembled rigid rods on triangular lattices is investigated using 
Monte Carlo simulation. 
We find a continuous transition between an ordered phase, where the rods are oriented along 
one of the three (equivalent) lattice directions, and a disordered one. 
We conclude that equilibrium polydispersity of the rod lengths does not affect the critical behavior, 
as we found that the criticality is the same as that of monodisperse rods on the same lattice, 
in contrast with the results of recently published work on similar models.
\end{abstract}
\maketitle

Recently experimentalists have acquired the ability to control the interactions 
between colloidal particles, with dimensions in the nanometer-to-micrometer range, 
providing new windows into the structural and thermodynamic behavior of colloidal suspensions. 
Of particular interest are the so-called "patchy colloids", 
the surfaces of which are patterned so that they attract each other via discrete bonding sites 
(patches) of tunable number, size and strength. 
Some of the collective properties of patchy colloids are being intensively studied with theory and simulations 
of primitive models and a number of results have been obtained \cite{SciortinoReview2010}. 

In systems with two bonding sites per particle, only (polydisperse) linear chains form 
and there is no liquid-vapor phase transition \cite{SciortinoChains2007}. 
If the linear chains are sufficiently stiff, however, they may undergo an ordering transition, 
at fixed concentration, as the temperature decreases below the bonding temperature. 
The description of self-assembled rods has to consider not only the effects of equilibrium
polydispersity but also the polymerization process. In this context, we proposed a model of 
self-assembled rigid rods (SARR), composed of monomers with two bonding sites that polymerize 
reversibly into polydisperse chains \cite{Tavares2009a} and carried out extensive Monte Carlo (MC) 
simulations to investigate the nature of the ordering transition of the model on the square 
lattice \cite{Almarza2010}. The polydisperse rods undergo a continuous ordering transition 
that was found to be in the 2D Potts q=2 (Ising) universality class, as in similar models 
where the rods are monodisperse \cite{Matoz2008b}. This finding refutes the claim that equilibrium 
polydispersity changes the nature of the ordering transition of rigid rods on the square lattice
from Ising to the Potts q=1 (percolation) universality class and questions a more recent one for a 
similar model on triangular lattices \cite{RamirezPastor2009,RamirezPastor2010a}. 

\begin{figure}
\includegraphics[width=75mm,clip=]{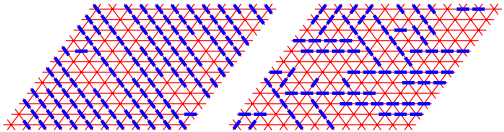}
\caption{Examples of ordered (left) and disordered (right) configurations
for the SARR model on triangular lattices. Monomers are represented with thick
segments lying on the lattice sites. Two nearest-neighbor monomers interact (and form
a bond) if the corresponding segments are in a head-to-tail configuration. }
\label{Fig1}
\end{figure}

In this note we (re)examine the original model of SARR \cite{Tavares2009a,Almarza2010} on triangular 
lattices (TL) using MC simulations. 
In the model a site is either empty or occupied by one monomer. 
Each monomer has two interacting patches pointing in opposite directions, $\pm {\bf s}$. 
The orientation (state) of the monomers, defined by the direction of the bonding patches, is restricted to 
the (three) lattice directions. The interaction potential can be described as follows:
Provided that two particles, $i$ and $j$, occupy
nearest-neighbor (NN) sites and provided that they are in the same state, the
energy is lowered by an amount $\epsilon$ only if their orientations are fully aligned with the
lattice vector ${\bf r}_{ij}$
(See Fig. \ref{Fig1}).
The bonding energy favors the self-assembly of rod-like lattice polymers (straight chains). 

The Grand canonical Hamiltonian of the system is given by:
\begin{equation}
H = - \epsilon \sum_{i=1}^M \sum_{k=1}^{p} f (| {\bf s}({\bf r}_i) \cdot {\hat \alpha}_k|)
f ( |{\bf s}({\bf r}_i+{\hat \alpha}_k  ) \cdot {\hat \alpha}_k|) - \mu \sum_{i=1}^M |{\bf s}({\bf r}_i)|,
\label{hamiltonian}
\end{equation}
where $i$ labels a lattice site,
${\hat \alpha}_k; k=1,\cdots, p$,
are unit vectors along the $p$ lattice directions ($p=3$ for TL);
${\bf s}({\bf r}_i)$ denotes the orientation at a given lattice site:
${\bf s}({\bf r}_i) = {\bf 0}$ for an empty site, while for
occupied sites ${\bf s}({\bf r}_i)$ is equal to one of ${\hat \alpha}_k$ vectors;
$M$ is the total number of sites;
$f(x)=1$ if $x=1$, and zero otherwise; and $\mu$ is the chemical potential.

An ordering transition will occur as the average rod length increases. 
In the ordered phase the rods will align preferentially along one of the $p$ lattice directions (See Fig. 1). 
In this  model, the only attractions between pairs of NN monomers are bonding ones. 
Additional {\it lateral} interactions that promote the condensation of monomers, 
leading to a competition between ordering of SARR and monomer condensation, are not considered, 
in line with the original SARR model \cite{Tavares2009a} on the square lattice 
\cite{Almarza2010,RamirezPastor2009}. The presence of only two bonding patches
and the absence of lateral interactions strongly suggest that the present model does not exhibit 
a discontinuous liquid-vapor transition at low temperatures \cite{SciortinoReview2010,SciortinoChains2007,Rouault}. 

The investigation of the ordering transition of SARRs on the TL is carried out through the analysis of the 
order parameter \cite{RamirezPastor2010a},
\begin{equation}
\delta = \frac{|\sum_{k=1}^{3} N_k {\hat \alpha}_k| }{ \sum_{k=1}^3 N_k },
\label{opq3}
\end{equation}
where $N_k$ is the number of monomers with orientation ${\hat \alpha}_k$.

In the Grand Canonical Ensemble, at a fixed chemical potential, the critical temperature, 
$T_c=T_c(\mu)$ is found by extrapolating to the thermodynamic limit the finite-size pseudo 
critical temperatures, $T_c(L)$, which may be defined in various ways ($L$ being the length
of the rhombic simulation box; $M=L^2$). 
Given the symmetry of the model, one can use the fourth-order cumulant of the order parameter 
distribution \cite{Landau_Binder,Weber}, $g_4=<\delta^4>/<\delta^2>^2 $, to define $T_c(L)$ as the 
temperature where the finite-size cumulant $g_4(T)$ takes the universal value, $g_{4c}$, 
for a given universality class and boundary conditions \cite{Landau_Binder}. We assume that 
the criticality of polydisperse rods on the TL is the same as that 
of monodisperse rods on the same lattice, i.e., Potts $q=3$ \cite{Vink,Matoz2008a}. We emphasize 
that this assumption is made for (computational) convenience and does not constrain the determination 
of the critical behavior, as discussed below. 

Although the value of $g_{4c}$ for the Potts q=2 model on the square lattice is well 
known \cite{Salas,Kamienarz} we have not found in the literature reliable estimates 
of $g_{4c}$ for the Potts $q=3$ model on TL with periodic boundary conditions, rhombic boxes, 
and order parameters defined as in Eq.(\ref{opq3}). 
We have therefore estimated its value by running simulations of the Potts $q=3$ model 
at the critical temperature\cite{Wu}, with the same box shape and boundary conditions,  
using the Swendsen-Wang algorithm \cite{Swendsen}. Different system sizes in the range 
$12 \le L \le 96$ were considered. The results were fitted to the scaling equation \cite{Bloete},
\begin{equation}
g_4(L,T_c) = g_{4c} + a L^{y_i},
\end{equation}
where $g_{4c}$, $a$, and $y_i$ are obtained from fits of the simulation results
or, alternatively, $y_i$ (the critical exponent associated to the so-called irrelevant 
field) is set to the theoretical value $y_i=-4/5$ \cite{Nienhuis,Shchur}. 
In the first case we find $y_i =-0.74 \pm 0.10$ and $g_{4c} = 1.168 \pm 0.002$; 
while setting $y_i=-4/5$ leads to $g_{4c}= 1.167 \pm 0.001 $. We have used the latter 
values in the finite-size scaling analysis reported below. 

We carried out {\it coupled} Grand Canonical Ensemble MC simulations. 
For a fixed value of $\mu$ several values of the temperature, 
$T_i=T_0+i\Delta T$ (with $i=0,1,\cdots,N_T-1$), are sampled in a single MC run using a 
simulation tempering algorithm \cite{Zhang2007}. 
This is achieved using a probability function given by:
\begin{equation}
P({\bf S}^M, T_i) =  \Omega( T_i ) \exp \left[ - H({\bf S}^M)/k_B T_i \right].
\label{PST}
\end{equation}
In order to obtain good sampling over all temperatures one has to use an appropriate weight 
function $\Omega(T_i)$. This was computed through an equilibration procedure following the 
usual strategies of the Wang-Landau-type algorithms\cite{Zhang2007,Wang,Lomba}. This simulation
tempering algorithm is known to enhance the sampling efficiency\cite{Zhang2008}.

\begin{figure}
\includegraphics[width=75mm,clip=]{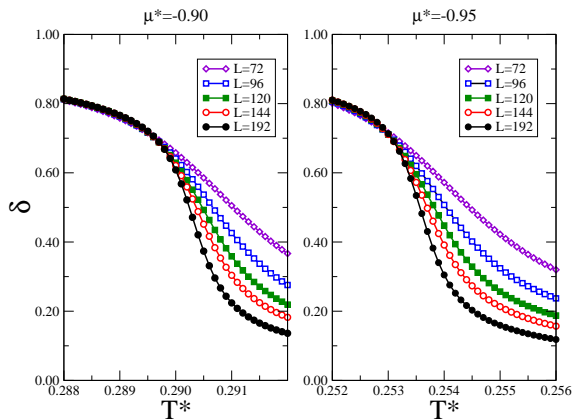}
\caption{Results for the order parameter $\delta$ as a function of 
the reduced temperature for different system sizes (as indicated in the legends);
 left panel $\mu/\epsilon=-0.90$, right panel $\mu/\epsilon=-0.95$.}
\label{Fig2}
\end{figure}

After preliminary runs to locate the critical region we run long simulations using typically 
between $N_T=20$ and $N_T=40$ values of $T_i$ around the
critical temperature. At each $\mu$ we considered different system sizes. As the interactions 
are restricted to NN, the lattices are 
split into three sublattices, where the sites do not interact energetically.
Simulation runs are organized in {\it sweeps}. In a sweep we update the state of every site 
and attempt one temperature change.
This is done by considering sequentially the three sublattices; we select for each site a 
{\it new} state ($k=0,1,2,3$)
(with $k=0$ denoting an empty site) with probabilities depending on the interaction energy, 
the value of $\mu$ and of the current temperature.
After updating all sites we attempt a temperature change by choosing 
at random (with equal probabilities) increasing or decreasing
the current temperature by an amount $\Delta T$, and accept or reject the change by considering 
the probability given by Eq. (\ref{PST}), and the usual Metropolis criterion \cite{Landau_Binder}. 
The length of a simulation run was $2 \times 10^8$ sweeps, and the results were split
into twenty blocks of $10^7$ sweeps for subsequent error analysis.
In Figure \ref{Fig2} we illustrate the results for the order parameter close to the transition
temperature, at two values of the chemical potential.

\begin{figure}
\includegraphics[width=75mm,clip=]{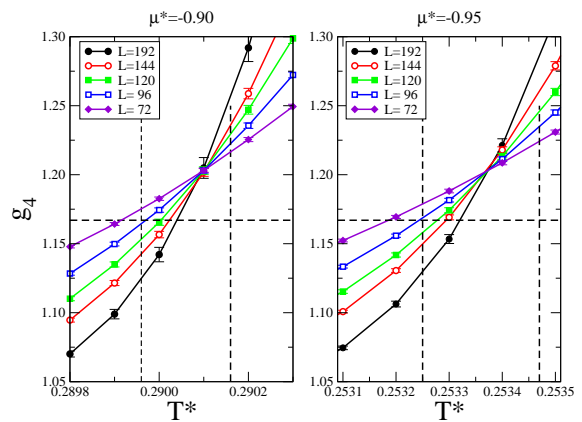}
\caption{Fourth order Binder cumulant at constant $\mu$ as a function of $T$, 
for different system sizes; left panel $\mu/\epsilon=-0.90$, right panel $\mu/\epsilon=-0.95$. Horizontal lines
depict the estimate of $g_{4c}$  for the two dimensional Potts q=3 universality class. Vertical
lines delimit the unbiased estimates of the critical temperature (See the text for the details).}
\label{Fig3}
\end{figure}

System-size dependent pseudo-critical temperatures, $T_{cL} = T_c (L)$ are computed by the 
matching criterion \cite{Wilding},
\begin{equation}
g_4(L,T_{cL}, \mu) = g_{4c},
\label{matching}
\end{equation}
where for convenience we set $g_{4c}$ to the universal value of the Potts $q=3$ universality class. 
Critical temperatures, $T_c$, were extrapolated by fitting the values $T_c(L)$ to scaling equations 
of the form,
\begin{equation}
T_c(L) = T_c + a L^{-b}.
\label{tcl}
\end{equation}
In order to avoid biasing the analysis we used two values for the exponent 
$b$: $b=(1+\theta)/\nu$
\cite{Wilding,Bloete} 
with $\theta=-y_i \nu$ and $\nu=5/6$ \cite{Wu} for 
Potts $q=3$ scaling, 
and $b=(1/\nu)_{q=1}=0.75$ for Potts $q=1$ \cite{Wu}; $\nu$ and $\theta$ are respectively
the correlation length and Wegner's correction to scaling exponents. 
However, the two values of $T_c$ were found 
to be very close.
The critical temperatures $T_c$ collected in Table \ref{table1} are those computed using 
the Potts $q=3$ scaling, which are consistent with the 
temperatures where the Binder cumulants cross (see Figure \ref{Fig3}). Notice that the crossing of
the $g_4(T)$ curves for different values of $L$ deviates slightly from the computed value of $g_{4c}$.
In order to constrain as little as possible the analysis of the criticality of SARRs on the TL we have 
computed {\it secondary} error bars (shown between curly brackets in Table \ref{table1}) that include 
the estimates for the critical temperature found using the Potts $q=1$ scaling.
\begin{table}[ht]
\begin{tabular}{lccc}
\hline \hline
System &  $\mu/\epsilon=-0.95$   &$\mu/\epsilon=-0.90$ & $\mu \rightarrow \infty$($\rho=1$) \\
\hline
n &  11 & 13 &  9 \\
$L_{min}$--$L_{max}$ & 84-192 & 72-192 & 60--144\\
\hline
$T_c^*$  &   0.25336(4)\{11\} & 0.29006(4)\{10\} &  0.47637(4)\{19\} \\
$\rho_c^*$ & 0.597(3)\{9\} & 0.688(3)\{6\} & --\\
\hline 
 $\beta/\nu$ &    0.110(22)\{57\} & 0.115(18)\{43\} & 0.126(9)\{31\} \\
$\gamma/\nu$ & 1.69(7)\{19\} & 1.72(5)\{12\} &1.70(4)\{12\} \\
 $1/\nu$ &   1.27(8)\{20\} & 1.28(6)\{12\} & 1.21(4)\{12\}  \\
 $(\alpha/\nu)$ &  0.40(28)\{38\}&  0.43(25)\{33\} & 0.45(28)\{32\}\\
\hline \hline
\end{tabular}
\caption{Finite-size scaling results from simulation: $n$ is the number of system
sizes used to compute critical properties and effective critical exponents.
$L_{min}$ and $L_{max}$ are the minimum and maximum system sizes used in
the finite-size scaling analysis. The results shown for $\alpha/\nu$  were
computed from the scaling of $(\partial \rho/\partial T)_{\mu}$, except for the
full lattice case ($\rho=1$) where $(\partial (H/M)/\partial T)$ was used. For finite
$\mu$ similar values of $\alpha/\nu$ were obtained using $(\partial \rho/\partial \mu)_T$ or
$(\partial (H/M)/\partial T)_{\mu}$. The critical exponent ratios for Potts $q=1$
universality class are $\beta/\nu=5/48 \simeq 0.104$, $\gamma/\nu=43/24\simeq 1.792$,
$1/\nu= 3/4$ and $\alpha/\nu=-1/2$. The corresponding values for Potts $q=3$ universality class
are $\beta/\nu=2/15 \simeq 0.133$, $\gamma/\nu =26/15 \simeq 1.733$, $1/\nu=6/5$, 
and $\alpha/\nu = 2/5$\cite{Wu}. Error bars are given in parentheses (curly brackets) 
in units of the last digit of the corresponding quantity.}
\label{table1}
\end{table}

\begin{figure}
\includegraphics[width=75mm,clip=]{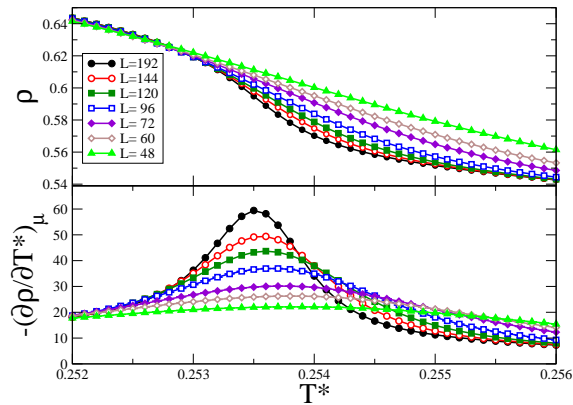}
\caption{Density and derivative of the density with respect to
the temperature at constant chemical potential $\mu/\epsilon=-0.95$.}
\label{Fig4}
\end{figure}

The critical behavior of the model was investigated by analyzing the system-size dependence of various properties 
at the extrapolated critical temperature. 
We fit the simulation results for a given property at $T_c$ to the expected scaling relation \cite{Landau_Binder}
$\delta(L) \propto L^{-\beta/\nu}$, $\chi(L) \propto L^{\gamma/\nu}$, and
$( \partial \ln <\delta(L)> /\partial T)_{\mu} \propto L^{1/\nu}$,
where  $\chi(L) = L^2 \left[ < \delta^2 > - < \delta>^2 \right]/k_B T$. 
$\beta$ and $\gamma$ are the
critical exponents for the order parameter and the susceptibility, respectively.
In addition, the quantities proportional to the second derivatives of 
the Grand Potential per unit volume with respect to the 
temperature and/or chemical potential, $(\partial (H/M) / \partial T)_{\mu}$, 
$(\partial \rho / \partial T)_{\mu}$, and $(\partial \rho / \partial \mu)_T$, 
are fitted to non-linear equations of the form,
\begin{equation}
\left(  \frac {\partial \rho } {\partial T}\right)_{\mu} = a_0 + a_1 L^{\alpha/\nu},
\end{equation}
where $\rho$ is the density (fraction of occupied sites), and
$\alpha$ is the specific heat critical exponent. In Fig. \ref{Fig4} we illustrate
the $\rho$, and of its derivative
$(\partial \rho/\partial T)_{\mu}$ at $\mu/\epsilon=-0.95$ around the critical temperature.
 In Table \ref{table1} we collect the estimates for the different critical exponents
(or exponent ratios). The uncertainty in the estimate of $T_c$ was taken into account and 
as we did for the critical temperatures, two estimates of the error bars are given, with the 
second one corresponding to error bars that are sufficiently large to include the critical 
temperature found using the Potts $q=1$ scaling.

For completeness we have computed the critical densities, by fitting the results to \cite{Wilding}: 
\begin{equation}
\rho_c(L,\mu,T_{cL}) = \rho_c(\mu) + a L^{-2+1/\nu},
\end{equation}
using the value of $\nu$ for the Potts $q=3$ universality class.

The results in table \ref{table1} clearly indicate that the critical behavior of the SARR model 
on the TL is much better described within the Potts $q=3$ than within the $q=1$ universality
class. Given that the former critical behavior was observed in the monodisperse case, we conclude that 
equilibrium polydispersity does not affect the critical behavior of rigid rod models, in contrast with the 
conclusion of Lopez et al. \cite{RamirezPastor2010a}. 
Even considering the largest error bars on the critical temperature, the values of the effective 
exponents $\nu$, and $\alpha/\nu$ are not compatible with Potts $q=1$ critical behavior.
The deviations observed in the crossings of $g_4(T)$ for different system sizes from the estimated value of 
$g_{4c}$ is most likely due to the importance of scaling corrections (low absolute value of $y_i$).

In previously published work \cite{Almarza2010} we discussed the reasons for the apparent $q=1$ critical behavior 
observed by Lopez et al. on the square lattice \cite{RamirezPastor2009}. 
The apparent $q=1$ behavior observed by the same authors on the TL \cite{RamirezPastor2010a} results also from using 
the density as the scaling variable. In fact, a simple but revealing analysis by Fisher\cite{Fisher}, shows that fixing 
the density in models such as those discussed here, corresponds to introducing a constraint that renormalizes the 
critical exponents. For the Potts $q=3$ universality class the renormalized correlation length exponent $\nu_X$ is 
$\nu_X = \nu/(1-\alpha) = 5/4$, which is close to the value of $\nu$ for the $q=1$ universality class $\nu_{q=1}=4/3$, 
reported by Lopez et al. \cite{RamirezPastor2010a}.

\acknowledgments
NGA gratefully acknowledges the support from the Direcci\'on General de Investigaci\'on 
Cient\'{\i}fica  y T\'ecnica under Grants Nos. MAT2007-65711-C04-04 and FIS2010-15502, and from the
Direcci\'on General de Universidades e Investigaci\'on de la Comunidad
de Madrid under Grant No. S2009/ESP-1691 and Program MODELICO-CM. MMTG and JMT acknowledge 
financial support from the Portuguese Foundation for Science and Technology (FCT) under 
Contracts nos.\ POCTI/ISFL/2/618 and PTDC/FIS/098254/2008.

\end{document}